# Wave Packet Dynamics in a Biased Finite-Length Superlattice

Herbert Kroemer[*)]

Department of Electrical and Computer Engineering,
and Department of Materials
University of California, Santa Barbara, CA 93106

## Abstract

In a superlattice containing a finite number of periods, the allowed values of the Bloch wave number form a discrete set, and the dynamics of an electron through $k$-space under the influence of an external force is necessarily that of a superposition wave packet composed of multiple Bloch waves. The present paper investigates this dynamics for a particularly simple class of "$k$-compact" wave packets, in which the spread over different $k$-values is minimized, and which propagate with an essentially rigid shape through $k$-space.

## 1) Introduction

This paper concerns itself with the question of how the electron dynamics in a biased superlattice changes relative to that in an infinitely long superlattice, as the number of cells is reduced. In the extreme limit of, say, one or two cells, it becomes of course meaningless to speak of a superlattice at all. But it is far less clear what happens for an only slightly larger number of cells, say, 10 cells. At what point does it become meaningful to speak of a superlattice?

The central issue is to what extent the dynamics can be described in terms of a dynamics in $k$-space. A familiar textbook theorem in solid-state physics states that, under the influence of an applied force $F$, the (reduced) Bloch wave number $k$ of an electron in a one-dimensional periodic potential changes with time according to "Newton's Law in $k$-Space,"

$$\hbar \frac{dk}{dt} = F, \qquad (1a)$$

with an obvious generalization to three dimensions. The following additional rule holds: Once $k$ reaches the edge of the reduced Brillouin zone, it re-enters the zone from the opposite edge (umklapp process).

The form (1a) treats the wave number $k$ as a continuous variable. However, in a potential of finite length, the allowed values of $k$ form a discrete set,

---

[*)] kroemer@ece.ucsb.edu

$$k_n = \frac{n}{N} \cdot \frac{2\pi}{a}, \text{ with } n = 0, 1, \ldots, N-1, \qquad (2)$$

where $N$ is the number of unit cells and $a$ the lattice period. Evidently, the (unsubscripted) wave number $k$ in (1a) cannot refer to a discrete single Bloch wave, but must represent a certain average over a wave-packet superposition of a set of Bloch waves. The purpose of the present paper is to study the nature of this average in some detail, and to determine how compact (in $k$-space) the corresponding wave packet can be, as a function of the number of cells contained in the superlattice.

In addition to the *formal* reasons given above, a finite-length superlattice also requires, on *physical* grounds, the consideration of wave packets rather than single Bloch waves: We are interested in the electron dynamics inside the *bulk* of a superlattice, without complications due to end effects at whatever bias terminals are present. But in this case we must restrict ourselves to wave packets that stay within the bulk during their Bloch oscillations. This calls not only for wave packets whose spatial extent is significantly smaller than the length of the superlattice, it also requires a restriction on the amplitude of their real-space oscillations, as illustrated in Fig. 1. This restriction, in turn, requires a bias voltage $V$ that is significantly larger than the voltage equivalent of the width $\mathscr{E}_B$ of the allowed band inside which the oscillation is taking place,

$$eV > \mathscr{E}_B. \qquad (3)$$

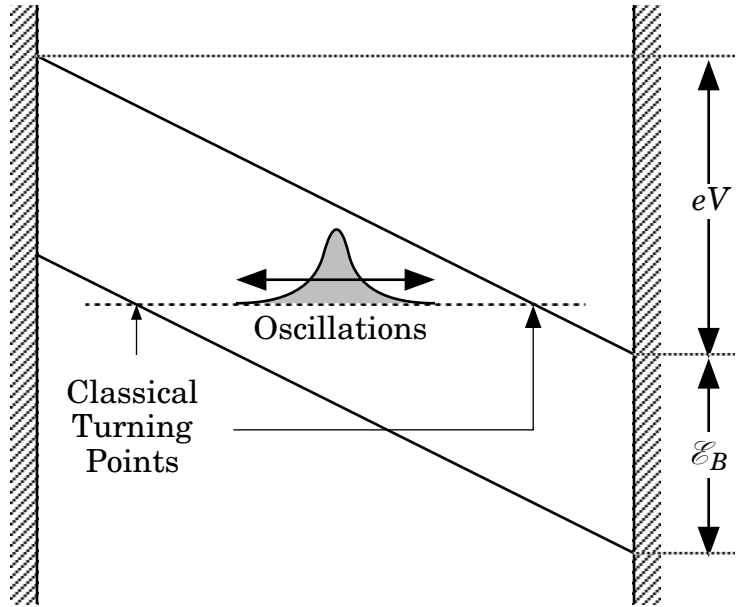

**Fig. 1.** Internal Bloch oscillations of a wave packet in a finite-length superlattice under bias.

These are of course the conditions under which the full $k$-space dynamics is of practical interest; hence these mathematical constraints are without *practical* importance.

To determine the exact nature of the average involved in (1a), we note first that all Bloch waves are eigenfunctions of the (unitary) lattice translation operator $\hat{T}$ with the (complex) eigenvalues $T_n = \exp(ik_n a)$, where $k_n$ is the Bloch wave number for the particular Bloch wave:

$$\hat{T}\psi(k_n, x) \equiv \psi(k_n, x+a) = e^{ik_n a} \cdot \psi(k_n, x). \tag{4}$$

As I have shown in a 1986 paper [1], what matters for the $k$-space dynamics is not the average of the $k_n$ values themselves, but that of the eigenvalues of the lattice translation operator,

$$\langle T \rangle = \langle e^{ik_n a} \rangle \equiv \sum_n P_n(t) \cdot e^{ik_n a}, \tag{5}$$

where $P_n$ is the (time-dependent) occupation probability for the state $|n\rangle$. More specifically, $\langle T \rangle$ must obey the relation

$$\hbar \frac{d}{dt}\langle T \rangle = iFa \cdot \langle T \rangle. \tag{6}$$

This relation remains valid for a superlattice of finite length, provided the wave function of the wave packet vanishes at the ends of the superlattice, as implied in Fig. 1. We will elaborate on this restriction in Sec. 6.

The solutions of (6) are of the form

$$\langle T \rangle = R \cdot \exp[iK(t) \cdot a], \tag{7}$$

where $R$ is a constant, and the quantity $K$ is time-dependent and obeys

$$\hbar \frac{dK}{dt} = F. \tag{1b}$$

Except for notation, this is of course nothing other than (1a). Evidently, it is this quantity $K$, defined by (7), that serves as the average for which (1a) is an exact law. Without loss in generality, we may pick the zero of the time $t$ such that $R$ is real and positive.

The form (7) automatically includes umklapp processes. Furthermore, it remains valid even in the presence of transitions between bands, i. e., Zener tunneling.

The quantity $\langle T \rangle$ evidently moves with a uniform rate of rotation along a circular trajectory of radius $R$ in the complex plane, with a rotation period

$$t_B = h/Fa = 2\pi/\omega, \tag{8}$$

where

$$\omega = Fa/\hbar \tag{9}$$

is the (angular) Bloch frequency.

For a potential containing $N$ unit cells, there will be $N$ allowed values of $\exp(ik_n a)$, equidistantly distributed around the unit circle of the complex plane, with $T_0 = 1$ being one of the allowed values.

In the infinite-crystal limit of a single Bloch wave we have $R = 1$, but for a superposition of Bloch waves the quantity $\langle T \rangle$ necessarily falls inside the unit circle of the complex plane, implying $R < 1$. One of the purposes of the present study is a determination of how $R$ decreases with decreasing number of cells in the periodic potential.

## 2) Rigid-Packet Approximation

Superimposed on the uniform rotation of the *center* of the wave packet along a circle in the $T$-plane, there will be periodic "breathing" oscillations of the *shape* of the wave packet in $k$-space, at the harmonics of the Bloch frequency. The situation is somewhat analogous to the familiar textbook case of the oscillation of wave packets in a harmonic-oscillator potential. In that case, it is well known that, regardless of the nature of the superposition forming the wave packet, the expectation values of position and momentum oscillate purely sinusoidally with the natural frequency of the oscillator, with no higher harmonics. However, in general, the *shape* of the wave packet, expressed in terms of its probability density, will exhibit synchronous *internal* "breathing" oscillations at harmonics of the natural frequency. In the HO case, there *do* exist special superpositions in which the probability density of the wave packet has the form of a Gaussian identical to the HO ground state Gaussian, except that it oscillates rigidly in space, without internal breathing oscillations.

In our case of a superlattice, with its much more complicated potential, such special rigidly oscillating packets do not exist. The investigation of the internal breathing oscillations is a complicated problem; the details depend on the $n$-dependence of certain matrix elements involving Bloch waves with different $n$ (see, for example, Houston [2]). However, they are of less interest than the motion of the center of the wave packet. A good understanding of the latter can be obtained by making what I would like to call the *Rigid-Packet Approximation*, in which the internal breathing oscillations are simply neglected.

Consider the occupation probability $P_0(t)$ of the Bloch wave with $n = 0$. In the absence of interband transitions, it must be a periodic function of time, with a time-averaged norm of 1/*N*. By a suitable choice of the origin of the time scale, we may expand $P_0(t)$, without loss of generality, as a Fourier series containing only cosine terms,

$$P_0(t) = \frac{1}{N} \cdot [1 + A_1 \cos(\omega t) + A_2 \cos(2\omega t) + \ldots]$$
$$= \frac{1}{N} \cdot \left[1 + \sum_{m=1}^{M} A_m \cos(m\omega t)\right], \quad (10)$$

where *M* is the highest harmonic order included. We leave its value open for now.

Under the rigid-packet approximation, the occupation probabilities of the remaining Bloch waves will differ from that for $n = 0$ by simple time shifts by $nt_B/N = 2\pi n/\omega N$, without a change in the *shape* of the probability densities,

$$P_n(t) = \frac{1}{N} \cdot \left\{1 + \sum_{m=1}^{M} A_m \cos[m(\omega t - 2\pi n/N)]\right\}. \quad (11)$$

The coefficients $A_m$ cannot be chosen independently, but are subject to the constraint that the quantity $\langle T \rangle$ defined in (5) must have a time dependence proportional to $\exp(i\omega t)$, with no other terms occurring:

$$\langle T \rangle \equiv \langle e^{ik_n a} \rangle \equiv \sum_{n=0}^{N-1} P_n(t) e^{ik_n a}$$
$$= \sum_{n=1}^{N-1} \left\{\sum_{m=1}^{M} A_m \cos[m(\omega t - 2\pi n/N)] \cdot e^{ik_n a}\right\} = R \cdot e^{i\omega t}. \quad (12)$$

If we insert $k_n$ from (2), we can re-express $\langle T \rangle$ in the form

$$\langle T \rangle = \frac{1}{2N} \sum_{m=1}^{M} A_m e^{im\omega t} \cdot \sum_{n=0}^{N-1} \exp[-2\pi i(m-1)n/N]$$
$$+ \frac{1}{2N} \sum_{m=1}^{M} A_m e^{-im\omega t} \cdot \sum_{n=0}^{N-1} \exp[+2\pi i(m+1)n/N]. \quad (13)$$

To satisfy (12), all terms must vanish except the $m = 1$ term in the first line.

Consider first the second line, where this requirement leads to the condition

$$\sum_{n=0}^{N-1} \exp[+2\pi i(m+1)n/N] = 0. \quad (14)$$

This is automatically satisfied unless $m + 1$ is an integer multiple of $N$. To eliminate such terms, we select the upper summation limit $M$ as

$$M = N - 2. \tag{15}$$

Inspection shows that in this case the terms with $m > 1$ in the first line also vanish, leaving us with the $m = 1$ term,

$$\langle T \rangle = \tfrac{1}{2} A_1 \cdot e^{i\omega t}. \tag{16}$$

Evidently

$$R = \tfrac{1}{2} A_1. \tag{17}$$

We shall see shortly that, for all but the shortest superlattices, the truncation condition (15) is a very "safe" one, in the sense that the highest included terms are already negligibly small.

Of course, the overall superposition must be normalized:

$$\sum_{n=0}^{N-1} P_n(t) = 1. \tag{18}$$

Inserting the probabilities specified by (11) shows that this condition is automatically satisfied if the Fourier series (11) contains no terms for which $m$ is an integer multiple of $N$, a possibility already excluded by our truncation condition (15).

### 3) A "*k*-Compact" Bloch Wave Packet

We are evidently free to choose $M$ coefficients $A_m$. We select them in a way to insure that the probability $P_0$ for Bloch wave with $n = 0$ is sharply peaked at $t = 0$ and at all multiples of a Bloch period, but goes to zero for $\omega t = \pi$ and at all odd multiples of $\pi$. Given the form (10), this condition leads to the requirement

$$-A_1 + A_2 - A_3 + \ldots (-1)^M A_M = \sum_{m=1}^{M} (-1)^m A_m = -1. \tag{19a}$$

We further require that not only $P_0$ itself vanishes for $\omega t = \pi$ and all odd multiples of $\pi$, but that its time-dependence is "maximally flat" in the vicinity of these minima, to insure a broad range of near-vanishing probability. We implement this idea by requesting that all even-order derivatives up to order $2(M-1)$ also vanish. This leads to the additional $M - 1$ conditions

$$\sum_{m=1}^{M} (-1)^m m^{2s} A_m = 0, \text{ for } s = 1, \ldots, M-1. \tag{19b}$$

The set (19a,b) is an inhomogeneous set of $M$ linear equations for the $M$ coefficients $A_n$. It is shown in Appendix A that the set has the solutions

$$A_m = \frac{2 \cdot (M!)^2}{(M+m)!(M-m)!}. \tag{20}$$

This implies

$$R = \tfrac{1}{2} A_1 = \frac{(M!)^2}{(M+1)!(M-1)!} = \frac{M}{M+1} = \frac{N-2}{N-1}, \tag{21}$$

provided $N > 1$. For $N = 2$, $R = 0$, for $N = 3$, $R = 0.5$, slowly increasing with increasing $N$. For $N = 10$, $R = 0.889$.

Fig. 2 shows the occupation probability of the state with $n = 0$, for $N = 10$, as a function of time.

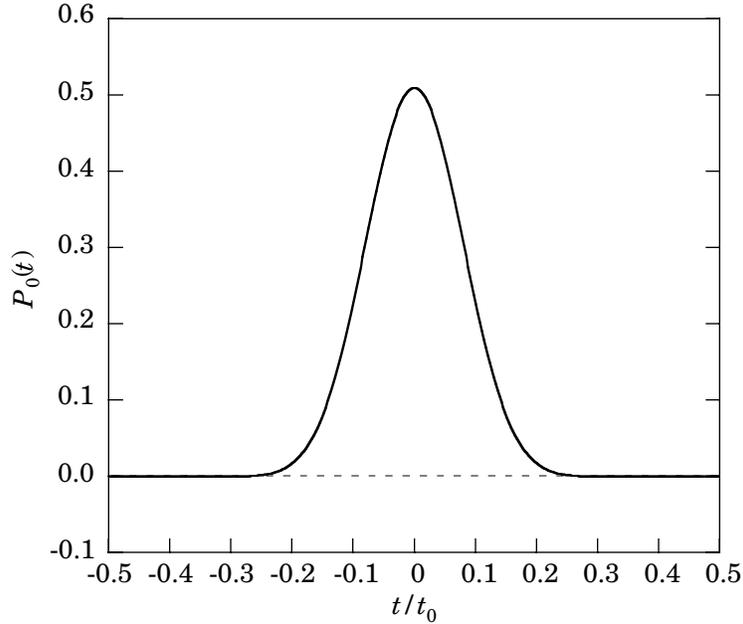

**Fig. 2.** Occupation probability for the Bloch wave with $n = 0$, as a function of time for one Bloch cycle, assuming a periodic potential with $N = 10$ unit cells (i. e., $M = 8$). At $|t/t_B| = 0.25$, the probability has dropped to 0.0019; it is essentially negligible beyond this point.

In the cases of greatest practical interest, the quantity $M$, while finite, will still be large compared to unity, and the dominating terms in the expansion will be those with $m \ll M$. The factorials in (20) may then be approximated by the Stirling approximation (which is accurate to 1% already for 8!). Working with the logarithms of $A_m$,

$$\ln A_m \approx \ln 2 - \left(M + \tfrac{1}{2}\right) \cdot \ln\left[1 - (m/M)^2\right] + m \cdot \ln \frac{1 - m/M}{1 + m/M} \qquad (22)$$
$$\approx \ln 2 - (m/M)^2 \cdot \left(M - \tfrac{1}{2}\right),$$

where the second line follows by expanding the logarithms for $m \ll M$. Exponentiating leads to

$$A_m \approx 2 \cdot \exp\left[-m^2 / (\Delta m)^2\right], \qquad (23\text{a})$$

where

$$\Delta m = M \Big/ \sqrt{M - \tfrac{1}{2}} = (N - 2) \Big/ \sqrt{N - \tfrac{5}{2}} = \sqrt{N} \cdot \left(1 - \tfrac{3}{4N} + \ldots\right). \qquad (23\text{b})$$

The dots refer to terms of order $1/N^2$ or higher, which can be neglected for all but the shortest superlattices. Throughout the rest of this paper, we will ignore terms of this and higher order. As an example, for $N = 10$ we have $\Delta m = 2.92$ for the exact form, 2.93 for the approximation. The following table shows, again as an example, the eight $A$-values for $N = 10$, both from (20) and from the approximation (23).

| $m$ | 1 | 2 | 3 | 4 | 5 | 6 | 7 | 8 |
|---|---|---|---|---|---|---|---|---|
| $A_m$ (Exact) | 1.778 | 1.244 | 0.679 | 0.283 | 0.0870 | 0.0186 | $2.49 \times 10^{-3}$ | $1.55 \times 10^{-4}$ |
| $A_m$ (Stirling) | 1.779 | 1.252 | 0.697 | 0.307 | 0.107 | 0.0294 | $6.42 \times 10^{-3}$ | $1.01 \times 10^{-3}$ |

Evidently, the terms fall off rapidly for $m > 3$, becoming negligibly small for $m = M$, thus justifying our neglect of higher-order terms beyond $M$. Also note that the Stirling approximation gives an excellent account of the dominant lower-order terms.

## 4) Width of the *k*-Compact State in *k*-Space and Real Space

In the foregoing, we determined the time-dependence of the occupation probability for the single Bloch wave $n = 0$. We now turn to the complementary question of the distribution of the probability over the whole set of Bloch waves at a given instant of time. Without significant loss in generality, we may pick $t = 0$. We obtain, from (11),

$$P_n(0) = \frac{1}{N} \cdot \left[ 1 + \sum_{m=1}^{M} A_m \cos(-2\pi mn/N) \right]$$
$$= \frac{1}{N} \sum_{m=-M}^{+M} \frac{A_m}{2} \cos(2\pi mn/N). \tag{24}$$

In the second line, we have extended the sum to $-M$, incorporating the "1" preceding the sum as the $m = 0$ term, with $A_0 = 2$, from (20).

In the limit of large values of $M$, we may insert the approximation (23a) for $A_n$ and approximate the sum by an integral. As is shown in Appendix B, this leads to

$$P_n(0) = \frac{1}{\sqrt{\pi}} \cdot \frac{1}{\Delta n} \cdot \exp\left[ -\left( \frac{n}{\Delta n} \right)^2 \right], \tag{25}$$

where

$$\Delta n = \frac{N}{\pi \cdot \Delta m} \approx \frac{1}{\pi} \sqrt{N} \cdot \left( 1 + \frac{3}{4N} \right). \tag{26}$$

Because of the relation (2) between $n$ and $k_n$, we are evidently dealing with a Gaussian distribution in $k$-space. If we make the transition from $t = 0$ to arbitrary values to time, the $k$-space probability distribution may be written

$$P_k(t) = \frac{1}{\sqrt{\pi}} \cdot \frac{1}{\Delta k} \cdot \exp\left\{ -\left[ \frac{k - K(t)}{\Delta k} \right]^2 \right\}. \tag{27}$$

Here $K(t)$ is the location of the peak of the Gaussian — which obeys the time evolution law (1b) — and

$$\Delta k = \Delta n \cdot \frac{\pi}{Na} \approx \frac{1}{a\sqrt{N}} \cdot \left( 1 + \frac{3}{4N} \right) \tag{28}$$

is the width in $k$-space.

In order to have a wave packet whose width $2\Delta k$ in $k$-space is small compared to the width $2\pi/a$ of the Brillouin zone, we must have

$$\pi \cdot \sqrt{N} \cdot \left(1 - \frac{3}{4N}\right) \gg 1, \tag{29}$$

a condition met for all but the shortest superlattices.

A Gaussian distribution in $k$-space implies a Gaussian distribution in real space, with a width

$$\Delta x = \frac{1}{2\Delta k} \approx \tfrac{1}{2} a \sqrt{N} \cdot \left(1 - \frac{3}{4N}\right), \tag{30}$$

large compared to a unit cell, but small compared to the overall length $Na$ of the periodic potential.

## 5) Real-Space Oscillations

The velocity in real space of the wave packet is the weighted average over the group velocities of the participating Bloch waves. In general, this average will be somewhat smaller in absolute magnitude than the velocity associated with the Bloch wave at the peak of the wave packet. To estimate the velocity reduction, we consider here only the simplest case:

(a) We assume that the number of cells $N$ is large enough that we may use the Gaussian approximation (27) and that the summation over $k$ may be replaced by an integration from $-\infty$ to $+\infty$:

$$v_P(t) \approx \int_{-\infty}^{+\infty} P_k(t) \cdot v(k) dk. \tag{31}$$

(b) We assume further that the group velocity of the individual Bloch waves is of the simple form

$$v(k) = \frac{1}{\hbar} \frac{d\mathscr{E}}{dk} = v_{\max} \cdot \sin ka, \tag{32}$$

where $v_{\max}$ is related to the width of the allowed band according to

$$v_{\max} = \frac{a\,\mathscr{E}_B}{2\hbar}. \tag{33}$$

Inserting (27) and (32) into (21), and executing the integration—using a procedure similar to that shown in Appendix B to derive (25)—leads to

$$v_P(t) \approx \alpha \cdot v_{\max} \cdot \sin[K(t)a], \tag{34}$$

with a reduction factor

$$\alpha = \exp\left[-\left(\tfrac{1}{2}\Delta k \cdot a\right)^2\right] \approx \exp\left[-\frac{1}{4N}\left(1+\frac{3}{2N}\right)\right]. \tag{35}$$

In the last form, we have replaced $\Delta k$ via (28). Evidently, for any reasonably large value of $N$, the velocity reduction is small (approximately 2.2% for $N = 10$). The effect of the reduction factor is the same as if the band width $\mathscr{E}_B$ had been reduced by $\alpha$.

The real-space amplitude of the wave packet oscillation associated with a velocity given by (34) is

$$x_{\max} = \alpha \frac{\hbar \cdot v_{\max}}{a \cdot dK/dt} = \frac{\alpha \mathscr{E}_B}{2F}, \tag{36}$$

where we have drawn on (1b) and on (33). This is exactly the amplitude one would associate with the distance between the two classical turning point shown in Fig. 1, with the band width $\mathscr{E}_B$ reduced by $\alpha$. Note that, for a sufficiently strong force $F$, the oscillation amplitude may drop below the width of the wave packet, or, more relevantly, the with of the wave packet may exceed the oscillation amplitude. This is simply a consequence of our having attempted to construct a wave packet with a narrow distribution in $k$-space, which implies a wide distribution in real space, in the direction towards a single Bloch wave, which would be distributed over the entire crystal.

## 6) The End Region Problem

In Sec. 1, we explicitly pointed out the need — on physics grounds — to restrict ourselves to wave packets which, during their dynamics, stay entirely within the finite length of the superlattice, without overlapping with any adjacent terminal regions. It is useful to show that this restriction does not need to be introduced as an additional restriction on top of the underlying mathematical formalism, but is already inherent in the formalism itself.

This formalism was based on the relation (6) for the time evolution of the expectation value $\langle T \rangle$ of the translation operator. The rigorous derivation of (6), given in [1], draws on the fact that the difference in the potential energies of adjacent cells *inside* the superlattice is a constant,

$$V(x+a) - V(x) = -Fa. \tag{37}$$

More specifically, the derivation requires that the expectation value of this potential difference, for any normalized wave packet, obeys the simple relation

$$\langle \psi |[V(x+a) - V(x)]|\psi\rangle = -Fa. \tag{38}$$

This is of course correct inside an infinitely long superlattice biased by a uniform force $F$, but at the ends of a finite-length superlattice, (37) will in general not be true, and as a result, the derivation of (6) breaks down — unless one restricts the admissible wave packets to those that vanish at the ends of the superlattice. This is what we have done here; the specific requirement (3) is a necessary but not sufficient condition for the validity of (6),

One might be tempted to try to circumvent this restriction by considering the finite-length superlattice as being embedded within an infinitely long superlattice inside which (37) remains valid everywhere, and by accounting for the finite length by imposing periodic boundary conditions on the wave functions over the length of the embedded finite-length superlattice. However, periodic boundary conditions are equivalent to the idea that a moving wave packet, upon leaving the superlattice at one end, simply re-enters it at the opposite end, with amplitudes and phases intact. But this is of course an un-physical idea we must reject.

An inclusion of end effects, just as a treatment going beyond the rigid-packet approximation, would be a complicated task that lies outside the scope of the present paper.

## Appendix A

The system determinant of the set (19a,b) of linear equations may be written as

$$D = (-1)^\nu \cdot \Delta(x_1, \ldots, x_M), \tag{A1}$$

where $\nu$ is the number of negative columns in the set of equations, and

$$\Delta(x_1, \ldots, x_M) = \begin{vmatrix} 1 & 1 & 1 & \ldots & 1 \\ x_1 & x_2 & x_3 & \ldots & x_M \\ x_1^2 & x_2^2 & x_3^2 & \ldots & x_M^2 \\ \vdots & \vdots & \vdots & \ddots & \vdots \\ x_1^{M-1} & x_2^{M-1} & x_3^{M-1} & \ldots & x_M^{M-1} \end{vmatrix}, \tag{A2}$$

with

$$x_l = l^2. \tag{A3}$$

The determinant $\Delta$ in (A2) is what is called a *Vandermonde* determinant, discussed in standard texts on linear algebra and/or matrix analysis. For arbitrary values of the $x_l$, the determinant evidently vanishes whenever two of the columns are equal to each other, and one finds easily that its value is simply the product of all $M \cdot (M-1)/2$ possible differences $(x_i - x_j)$, with $i > j$, and both $i$ and $j$ from the set $1 \ldots M$:

$$\Delta(x_1,\ldots,x_M) = \prod_{\substack{i,j=1 \\ i>j}}^{M} (x_i - x_j) = \prod_{\substack{i,j=1 \\ i>j}}^{M} (i+j) \cdot \prod_{\substack{i,j=1 \\ i>j}}^{M} (i-j). \tag{A4}$$

In the last equality we have drawn on (A3).

The solutions $A_m$ are the ratios of two determinants,

$$A_m = (-1)^{m-1} \cdot \frac{\Delta_m}{\Delta}, \tag{A5}$$

where $\Delta_m$ differs from $\Delta$ by replacing $x_m$ everywhere with zero, leaving only a term $+1$ at the top of column $m$. The pre-factor $(-1)^{m-1}$ in (A5) accounts for the change in sign when the "$-1$" on the right-hand side of (19a) is substituted for a "$+1$" in the top row of one of the even-numbered columns in $\Delta$.

Evidently, in the ratio (A5) all those factors cancel that do not contain $x_n$. We may therefore split off these factors in both determinants. The numerator is easiest; we may write it as

$$\begin{aligned}\Delta_m &= \left[ x_M \cdot \ldots \cdot x_{m+1} \cdot (-x_{m-1}) \cdot \ldots \cdot (-x_1) \right] \cdot R_m \\ &= (-1)^{m-1} \cdot \left( \frac{M!}{m} \right)^2 \cdot R_m,\end{aligned} \tag{A6}$$

where $R_m$ is the product of all remaining factors, those *not* involving $x_m$. In the second line the $x_l$ have been replaced by $l^2$, according to (A3). The denominator determinant $\Delta$ may be evaluated analogously. After some manipulation one finds

$$\Delta = \frac{1}{2m^2} \cdot (M+m)!(M-m)! \cdot R_m. \tag{A7}$$

Inserting (A6) and (A7) into (A5) leads to the claimed result (20).

## Appendix B

If we insert (23a) into the last sum in (24), we obtain

$$P_n(0) = \frac{1}{N} \sum_{m=-M}^{+M} \exp\left[-m^2/(\Delta m)^2\right] \cdot \cos[2\pi m n/N]$$

$$= \frac{1}{N} \sum_{m=-M}^{+M} \Re\left\{\exp\left[-\frac{m^2}{(\Delta m)^2} + 2\pi i \frac{mn}{N}\right]\right\} \quad \text{(B1)}$$

$$= \frac{1}{N} \exp\left[-\left(\frac{\pi n \Delta m}{N}\right)^2\right] \cdot \Re\left\{\sum_{m=-M}^{+M} \exp\left[-\left(\frac{m}{\Delta m} - i\pi\frac{n\Delta m}{N}\right)^2\right]\right\}.$$

For large values of *M*, the last sum may be approximated by an integral in the complex plane, with the integration path shifted to the real axis:

$$\sum_{m=-M}^{+M} \exp\left[-\left(\frac{m}{\Delta m} - i\pi\frac{n\Delta m}{N}\right)^2\right] \approx \int_{-\infty}^{+\infty} \exp\left[-\left(\frac{m}{\Delta m}\right)^2\right] dm = \sqrt{\pi}\cdot\Delta m. \quad \text{(B2)}$$

Hence,

$$P_n(0) = \frac{\sqrt{\pi}\cdot\Delta m}{N} \cdot \exp\left[-\left(\frac{\pi n \Delta m}{N}\right)^2\right]. \quad \text{(B3)}$$

Because of (26), this may be written in the form (25).

## References

[1] H. Kroemer, "On the Derivation of $\hbar \cdot dk/dt = F$, the $k$-Space Form of Newton's Law for Bloch Waves," *Am. J. Phys.*, vol. **54**, pp. 177-178 (1986). See also Appendix E in C. Kittel, *Introduction to Solid-State Physics*, Wiley, New York, 1986 and later.

[2] W. V. Houston, "Acceleration of Electrons in a Crystal Lattice," *Phys. Rev.*, vol. **57**, pp. 184-186 (1940).